\documentstyle[11pt,twoside,colloqOHP2010,epsfig]{article}
\begin{document}

\title{Tidal instability in exoplanetary systems evolution}
\author{D. C\'ebron$^1$, C. Moutou$^2$, M. Le Bars$^1$, P. Le Gal$^1$, R. Far\`es$^2$}

 \affil{$^1$ Institut de Recherche sur les Phenomenes Hors
 Equilibre, UMR 6594, CNRS-Aix-Marseille University,
  49 rue F.Joliot-Curie, 13384 Marseille Cedex 13, [cebron@irphe.univ-mrs.fr]}
\affil{ $^2$ Laboratoire d'Astrophysique de Marseille, UMR 6110,
  CNRS-Aix-Marseille University, 38 rue F. Joliot-Curie, Marseille}

\begin{abstract}
A new element is proposed to play a role in the evolution of
extrasolar planetary systems: the tidal (or elliptical) instability. It comes
from a parametric resonance and takes place in any rotating fluid
whose streamlines are (even slightly) elliptically deformed. Based
on theoretical, experimental and numerical works, we estimate the
growth rate of the instability for hot-jupiter systems, when the
rotation period of the star is known. We present the physical
process, its application to stars, and preliminary results obtained
on a few dozen systems, summarized in the form of a stability
diagram. Most of the systems are trapped in the so-called "forbidden zone",
where the instability cannot grow. In some systems, the tidal
instability is able to grow, at short timescales compared
to the system evolution. Implications are discussed in the framework
of mis-aligned transiting systems, as the rotational axis of the
star would be unstable in systems where this elliptical instability
grows.

\end{abstract}


\section{Introduction}

The role of tides in the evolution of systems composed of a star and a close-in companion has been investigated in several studies recently, in order to tentatively explain for instance: i) the spin-up of stars with hot jupiters (Pont 2009), ii) the radius anomaly of strongly irradiated planets (Leconte et al 2010), iii) the synchronisation or quasi-synchronisation of the stellar spin (Aigrain et al 2008)... Such studies always account for the static tides. Most recent observations have attracted our interest to add a new ingredient in this study, the tidal instability. The star $\tau$ Boo has a massive planet in short orbit (4.5 $M_{Jup}$ minimum mass and 3.31-day period), and the remarkable property of a stellar surface rotating synchronously with the planetary orbit. It is possible that tides from the planet onto the star have synchronised the thin convective zone of this F7 star, since the mass ratio between the planet and the convective zone is  more than 10. Recurrent spectropolarimetric observations have allowed to reconstruct the global magnetic topology of the star since 2006, and its evolution. Two polarity reversals have been observed in two years (Donati et al, 2008, Far\`es et al, 2009), an evidence for a magnetic cycle of 800 days, much shorter than the Sun's (22 yrs). The role of the planetary tides on the star in this short activity cycle was questioned; a strong shear may take place at the bottom of the convective zone, triggering a more active and rapidly evolving dynamo. 

The misalignement of one third of transiting hot jupiters (Winn, 2010) also questioned the role of tides in such systems, since the tides are responsible for alignement of the planes, circularization of the orbit and synchronisations of periods. The classical idea of planet formation and migration within a disk was also challenged by such observations. Tidal instabilities may cause the rotational axis of both bodies in the system to change orientation with time, at a relatively short timescale. Misaligned systems could thus show unstable rotation axes of stars, rather than tilted orbital planes of the planet. Tidal implications on the internal structure of planets were already reported (Leconte et al 2010). Since the power dissipated by tides could generate radius anomaly, then the additional power generated by tidal instabilities should also be taken into account in planet modeling.

In this article, we define the tidal instability and apply its properties to hot-jupiter systems.

\section{Tidal instability}

\subsection{Definition of the model and dimensionless parameters}
The tidal instability corresponds to the astrophysical version of the generic elliptical instability, which affects all rotating fluids with elliptically-deformed streamlines (Kerswell 2002). The origin of the instability is a resonance between inertial waves of
the rotating fluid and the tidal wave i.e.
the underlying strain field responsible for the elliptic deformation
(e.g. Waleffe, 1990). Recent studies of tidal
instabilities have been reported in the context of binary stars
(Rieutord, 2003 and Le Bars et al 2010) or planetary cores
(Aldridge et al 1997, C\'ebron et al 2010) where they could play an
important role in the induction of a magnetic field
(Kerswell \& Malkus 1998, Lacaze et al 2004, Herreman et al 2009) and even in
planetary dynamos 
(Malkus, 1993). The basic principles of the
tidal instability are recalled in the following.

Let us consider the star as a massive fluid body, with its own spin
and tidally deformed by an orbiting extra-solar planet. For our
purpose, this star can be modeled as a rotating flow inside a fixed
triaxial ellipsoidal shell of outer axes $(a,b,c)$, where $c$ is the
length of the polar axis. In the following, we will use the mean
equatorial radius $R_{eq}=(a+b)/2$ as the lengthscale. We introduce
also a timescale $\Omega^{-1}$ using the tangential velocity along
the deformed outer boundary at the equator $U=\Omega R_{eq}$. Then,
the hydrodynamic problem is fully described by five dimensionless
numbers: the ellipticity $\beta=|a^2-b^2|/(a^2+b^2)$, the shell aspect ratio $\eta$, the aspect ratio $c/b$, the Ekman number
$E=\nu/(\Omega R_{eq}^2)$, where $\nu$ is the kinematic viscosity of
the fluid and the ratio $T$ of the orbital period $T_{orb}=2 \pi/
\Omega_{orb}$ to the spin period $T_{spin}=2 \pi/ \Omega$. We will
also use the ratio between the orbital and the fluid angular
velocities $\Gamma=\Omega_{orb}/(\Omega_{spin}-\Omega_{orb})$.

\subsection{Calculation of the instability growth rate}
In most of the previous studies on the tidal instability, it is
assumed that the tidal deformation is fixed and that the excited
resonance is the so-called spin-over mode, which corresponds to a
solid body rotation around an axis perpendicular to the spin axis of
the system. This is indeed the only perfect resonance in spherical
geometry in the absence of rotation of the elliptical deformation
(Lacaze et al 2004). But in all natural configurations such as
binary stars, moon-planet systems or planet-star systems, orbital
motions are also present. This implies that the gravitational
interaction responsible for the tidal deformation is rotating with
an independent angular velocity $\Omega_{orb}$ different from the
spin of the considered star. This significantly changes the
conditions for resonance and the mode selection process, as recently
studied in Le Bars et al (2007, 2010)  and C\'ebron et al (2010). A very
important point is the apparition of a so-called 'forbidden zone'
where the tidal instability cannot grow whatever the deformation and
the rotation rates are. This forbidden zone is simply given by the
following kinematic conditions $-1 \leq T \leq 1/3$. This is a key point when observing natural systems possibly affected by the tidal instability. Outside from this forbidden zone,
the growth rate of the tidal instability is given by
\begin{equation}
\sigma=\left[ \frac{(3+2 \Gamma)^2}{16(1+\Gamma)^2}\ \beta-K\ E^{\gamma} \right]\left[1-\frac{1}{T} \right], \label{eq:sigmasimple}
\end{equation} \\
where the first term on the right-hand side comes from the inviscid
local analysis (Le Diz\`es 2000, le Bars et al. 2010). The second
term is due to the viscous damping of the instability. The
pre-factor $K$ depends on the excited mode of the tidal instability and on the aspect ratio $\eta$, the exponent $\gamma$ depends on the kind of the viscous
damping. If the surfacic damping is dominant, as in the case of
no-slip boundaries, $\gamma=1/2$. On the contrary, in the absence of
surfacic viscous damping, the volumic damping in the bulk of the
flow is dominant and then $\gamma=1$. In the case a planetary core
or experiments, the boundary conditions are no-slip and the
situation is unambiguous: the surfacic damping is dominant. But in
the case of stars, the boundary conditions between the radiative and
the convective zones are less clearly defined. On one hand, Tassoul
(1987, 1995) argue that dissipation scales as $E^{1/2}$ (see
also Tassoul \& Tassoul, 1992, 1997). On the other hand,
Rieutord (1992, 2007, 2008) argue that this dissipation is
rather proportionnal to $E$ (see also Rieutord \& Zahn, 1997). In
this work, we will consider both models in order to obtain
conservative results.

\subsection{Observables due to the tidal instability}

The tidal instability is expected to be dynamo capable, as already
suggested in Malkus (1993), Lacaze et al. (2006) or Arkani-Hamed et al (2008). In this case, the magnetic field dynamics should be very different from the dynamics driven by
a thermo-solutal convective dynamo. Another interesting feature from an astrophysical point of view is the inclination of the rotation axis of the flow due to the spinover
mode. Indeed, this could be directly related to the spin-orbit
obliquity angle. In this case, there is no need to evoke a
supplementary non-observable third body  to justify the obliquity.
The inclination of the rotation axis would be a direct consequence
of the tilted rotation axis of the stellar flow. Note that in this
case, as there is no supplementary torque acting on the star, the
conservation of the angular momentum will imply that the rotation
axis of the Hot Jupiter will also be tilted.

Finally, the tidal instability modifies the whole flow, which leads
to a stronger viscous dissipation. This supplementary dissipation is
larger than the simple dissipation induced by the tidal strain of
the elliptically distorted basic flow. This supplementary dissipation could
be very important for the synchronization process
(Le Bars et al 2010). We also expect that this extra dissipation
and supplementary heat release could be related to the observed radius anomaly.

\section{Application to hot-jupiter systems}
Using the previous theoretical results, we can calculate the
threshold of different exoplanetary stars and then draw a stability diagram.
Note that we consider that a body is stable (respectively
unstable) if the mean value of the growth rate over an orbit is
negative (respectively positive). In order to estimate the tidal ellipticity $\beta$, we consider a polytropic fluid (equation of state: $p=A\ \rho^{1+1/n}$ where $p$ is the pressure, $\rho$ the density and $(A,n)$ two constants) of index $n=3/2$ and calculate the static equilibrium tides (details will be given elsewhere).

\subsection{The stellar model}

\begin{figure}[h!]
\begin{center}
\epsfig{width=6cm,file=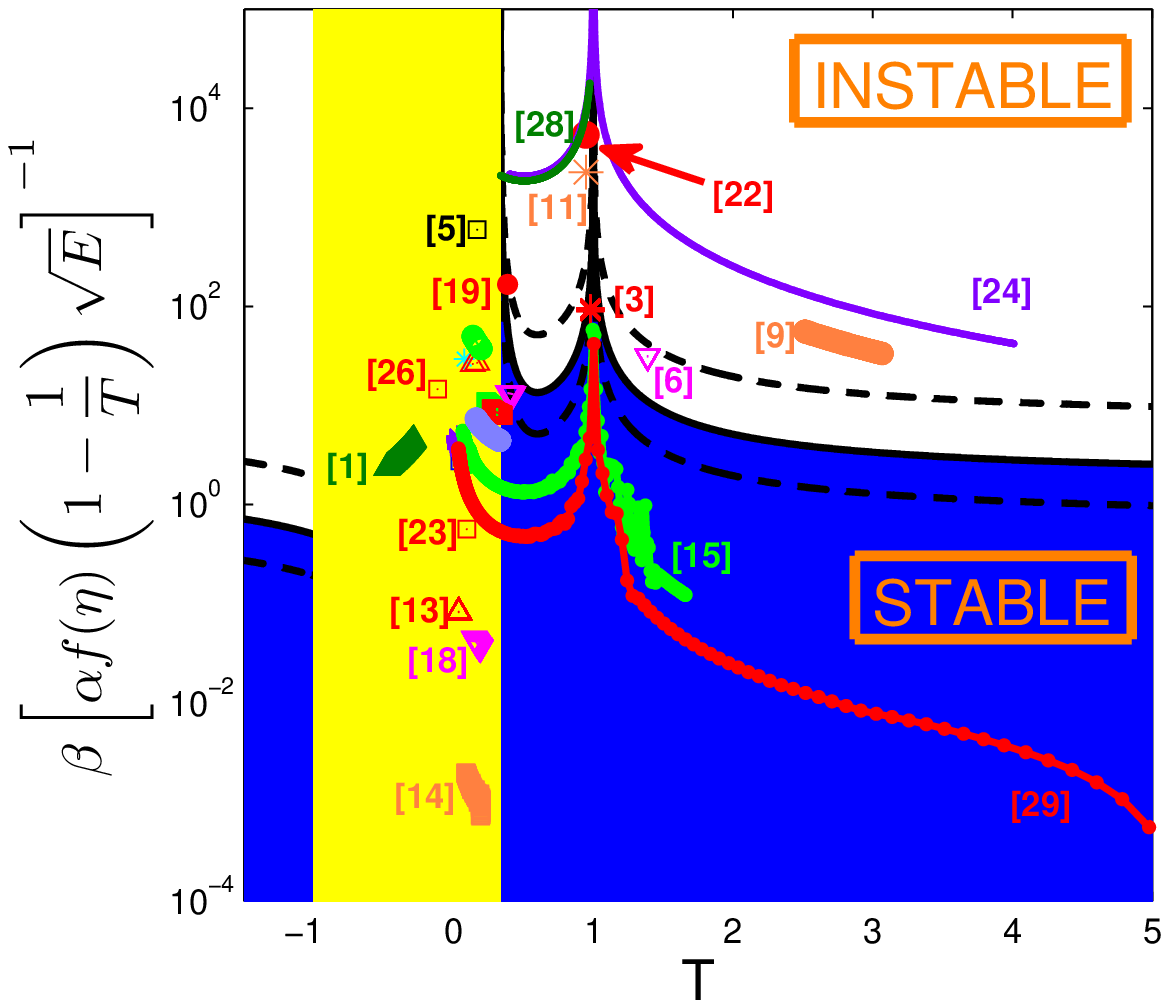} 
\epsfig{width=5.5cm,file=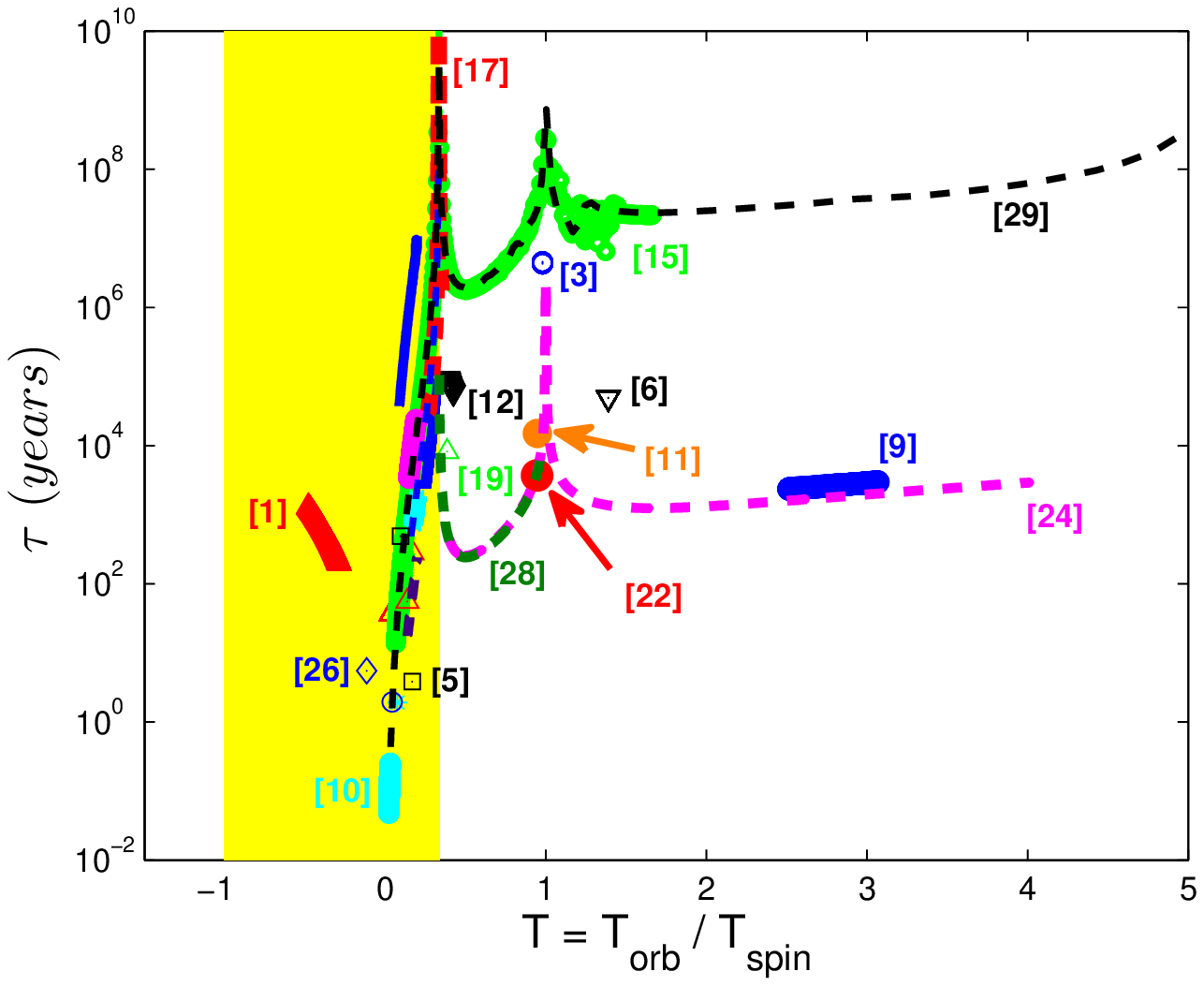}
\caption{(Left) Stability diagram with the considered stars using a dominant surfacic dissipation as in Tassoul's model. The blue zone is the stable zone, defined from equation (2) with $\alpha=2.62$, whereas the black dashed line represent respectively $\alpha=1$ and $\alpha=10$. The
yellow zone is the so-called 'forbidden zone' where no tidal
instability can appear. The numbers represent the stars of the
systems considered. Among the $29$ stars considered, $17$ are in the
forbidden zone and $8$ are unstable. (Right) Typical timescale $\tau=\overline{\sigma}^{-1}$ for the growth of the tidal instability in extra-solar stars, using a bulk dissipation as in Rieutord's model.} \label{cebronfig1}
\end{center}
\end{figure}

We model the considered stars as a two zone system: a convective
zone where thermal convection is the dominant physical process and a
radiative thermally stratified zone. In the cases considered here,
the convective zone is systematically above the radiative zone.
First, let us focus on the external convective zone, where the
ellipticity deduced from the equilibrium tides theory should not be
modified too much by the compressibility effects. Considering this
external layer with a ratio $\eta$ given by the so-called RCZ, we
estimate its inviscid growth rate using equation
(\ref{eq:sigmasimple}). Note that the convective fluid motions
should not significantly modify the growth rate as shown in
C\'ebron et al (2010b) and Lavorel \& Le Bars (2010). The molecular kinematic
viscosity inside this zone is calculated considering the molecular
kinematic viscosity of the Sun and assuming a temperature dependance
in $T^{-5/2}$. This leads (e.g. Rieutord 2008) to
$\nu=0.001\ (5800/\varsigma)^{5/2}$, with $\varsigma$ the surface
temperature of the star in Kelvin unit, and $\nu$ in $m^2\ s^{-1}$.

\subsection{Diagram of stability using Tassoul's model}

Considering that the viscous damping term in the growth rate
expression scales as $E^{1/2}$, the situation is similar to the
situation met in experiments, i.e. with no-slip boundaries. In this
case, we have $K=\alpha\ (1+\eta^4)/(1-\eta^5)\alpha\ f_{(\eta)}$, with $\alpha =
O(1)$ a constant, equal to $\alpha=2.62$ for the spinover mode of
the tidal instability (see e.g. Hollerbach \& Kerswell 1995, Lacaze et al 2005). This allows to define a threshold and then to plot the stability diagram shown in the figure \ref{cebronfig1}
which summarizes the results. The equation (\ref{eq:sigmasimple}) allows to write the stability condition $\sigma \geq 0$ equation under the form:\begin{eqnarray}
Y=\beta \left( \alpha f(\eta) \sqrt{E} \right)^{-1} \left(1- \frac{1}{T} \right)^{-1}  \geq \frac{16\ (1+\Gamma)^2 }{(2 \Gamma+3)^2 (1-1/T)}.
\end{eqnarray}
which defines the vertical axis $Y$ used in the figure \ref{cebronfig1}.

As can be seen on Fig. 1, among the $12$ stars outside the forbidden
zone, $8$ stars are found to be unstable ($\overline{\sigma}\ >0$).

\subsection{Diagram of stability using Rieutord's model}

In this case, the prefactor $K$ of the viscous damping term in
equation (\ref{eq:sigmasimple}) is simply the square of the wavenumber $k=2 \pi/
\lambda=\sqrt{K}$ of the excited mode. However, the value of $k$, needed to
calculate explicitly the damping term, changes with the excited
mode. A lower bound estimation for $k$ is given by $k=\pi/(1-\eta)$,
which corresponds to the first unstable mode (a half wavelength
inside the gap of the shell). A relevant upper bound is difficult to
obtain. However, the balance between the typical inviscid growth
rate $\beta$ and the volume damping term leads to a typical
wavelength $\lambda=2 \pi/k$ given by $l \sim \sqrt{E/\beta}$. Then, with
typical values $\beta/E \sim 10^{-10}$ ,
we obtain $\lambda \sim 10^{-5}$, which is larger than
the Kolmogorov microscale $E^{3/4} \sim 10^{-12}$ given by the
Kolmogorov theory which is the scale limit between molecular
dissipation and turbulent dissipation. This shows in particular that
the different scales are well separated in the flow, allowing the
elliptic, large scale instability to grow. Assuming $k \sim O(1)$, the calculated
dissipative term is so small that it can be neglected. We thus can
consider that this free slip boundaries case leads to a growth rate
of the same order of the inviscid growth rate $\sigma$. The
associated typical timescales $\tau=1/\overline{\sigma}$ is thus chosen to
represent the strength of the instability (see Figure 1). This model is correct provided that $\tau \gg \Omega_{orb}^{-1}$, which allows to consider $\overline{\sigma}$, and $\tau \ll \tau_2$ where $\tau_2$ is the typical time of stellar evolution. For instance, considering the synchronization time $\tau_2$, it follows that the presence of the tidal instability is expected in the range $10\ y \ll \tau \ll 10^6\ y$. This gives $8$ unstable stars, which are the same as in the previous section except that the star $[3]$ is replaced by the star $[12]$ (see the next section).

\section{Results and discussions}

Due to a selection bias (hot Jupiters have orbital periods less than 5-10 days and exoplanets are mainly found around low rotator stars with rotational periods more than 20 days, hence $T \leq 1/3$), 17 stars lie in the forbidden zone:
WASP-17, WASP-10, WASP-19, WASP-18, Kepler-5, TrES-3, HD41004B, CoRoT-7, GJ-674, HD189733, HAT-P-1, HD149026, TrES-2, TrES-1, CoRoT-1, HAT-P-7, WASP-14.
Outside the forbidden zone, the stars found to be unstable with both dissipation models are CoRoT-6 [6], $\tau$ Boo [11], CoRoT-2 [19], CoRoT-3 [22], XO-3 [28], HAT-P-2 [24]. Apart these ones, HD179949 [12], HD80606 [29], CoRoT-4 [3] could also be unstable (with a lower level of confidency) whereas HD17156 [15] is probably stable. 
Tidal unstability could be invoked to explain the misaligned spin-orbit axes of XO-3b, CoRoT-3b and HD 80606b. The misalignement of WASP-17b, CoRoT-1b, HAT-P-7b a,d WASP-14b (all included in our sample but in the forbidden zone) could be explained by an evolution of the system involving tidal unstability occurring before the stellar spin-down. 
Finally, the short-cycled star $\tau$ Boo has been demonstrated to be among the most unstable stars of our sample: not only tides from the planet have an impact on the system's evolution, but now we can also describe explicitly this process with a physical mechanism, the elliptical instability.

The consequences of tidal instability in hot-jupiter systems have to be deeper investigated: internal structure of stars and planets, misalignement or evolution of the stellar (and planetary) rotation axis, effects on the dynamo. The calculations have to be applied to a larger sample of systems, where the stellar rotation period would be measured with accuracy. Further results will be presented in a forthcoming paper by C\'ebron et al (2011).

\end{document}